\documentclass[a4paper,11pt]{article}
\pdfoutput=1 % if your are submitting a pdflatex (i.e. if you have
\usepackage{jcappub} % for details on the use of the package, please
\usepackage[T1]{fontenc} % if needed
\usepackage{subfigure}

\def \beq{\begin{equation}}
\def \eeq{\end{equation}}

\title{\boldmath Tachyon inflation in the large-$N$ formalism}

\author[a,b]{Nandinii Barbosa-Cendejas,}
\author[a]{Josue De-Santiago,}
\author[a]{Gabriel German,}
\author[a]{Juan Carlos Hidalgo,}
\author[a]{Refugio Rigel Mora-Luna.}

\affiliation[a]{Instituto de Ciencias F\'{\i}sicas, Universidad Nacional Aut\'onoma de M\'exico,\\Apdo. Postal 48-3, 62251 Cuernavaca, Morelos, M\'exico}
\affiliation[b]{Facultad de Ingenier\'ia El\'ectrica,
Universidad Michoacana de San Nicol\'as de Hidalgo,\\Morelia, Michoac\'an, M\'exico}

\emailAdd{nandini@fis.unam.mx}
\emailAdd{josue@fis.unam.mx}
\emailAdd{gabriel@fis.unam.mx}
\emailAdd{hidalgo@fis.unam.mx}
\emailAdd{rigel@fis.unam.mx}

\abstract{We study tachyon inflation within the large-$N$ formalism, which takes a prescription for the small Hubble flow slow--roll parameter $\epsilon_1$ as a function of the large number of $e$-folds $N$. This leads to a classification of models through their behaviour at large $N$. In addition to the perturbative $N$ class, we introduce the polynomial and exponential classes for the $\epsilon_1$ parameter. With this formalism we reconstruct a large number of potentials used previously in the literature for Tachyon Inflation. We also obtain new families of potentials form the polynomial class.  We characterize the realizations of Tachyon Inflation by computing the usual cosmological observables up to second order in the Hubble flow slow--roll parameters. This allows us to look at observable differences between tachyon and canonical single field inflation. The analysis of observables in light of the Planck 2015 data shows the viability of some of these models, mostly for certain realization of the polynomial and exponential classes.}

\begin{document}
\maketitle
\flushbottom

\section{Introduction}
\label{sec:intro}

From its inception, inflation  has been a extremely useful mechanism to address several issues which the old cosmology was unable to
explain \cite{inflation}. The inflationary paradigm has involved a
vast effort in model building and the variety of models is huge. 
 The nature of the inlfationary
field is not yet determined and in principle the tachyon can be
responsible for inflation. The tachyon field was brought up to
prominence by A. Sen \cite{sen:todas,Sen:2002an} who studied type II
string theory and the tachyon instability signals on D-branes. In
the context of the brane wold paradigm the tachyon field as also
been a subject of study (see for instance  \cite{German:2012rv}) and
references there in. The cosmological relevance of the tachyon was
first explored in \cite{Gibbons,cosh,Daniel}, where the expansion of the
universe was studied for various initial conditions (see also
\cite{Nozari:2013mba,Li:2013cem,tachyon1,tachyon2}). Independently of
its possible origin in string theory one can simply take the tachyon
field as another inflaton candidate and study its implications
without trying in a first approximation to understand its origin and
theoretical implications.

In the present article we study tachyon inflation in the so called large-$N$ formalism \cite{Boyanovsky:2005pw,Roest:2013fha,Garcia-Bellido:2014gna}, where relevant quantities are functions of the number of $e$-folds $N$, taken as an evolution variable instead of the usual inflaton field $\phi$, or cosmic time. The large-$N$ formalism has been successfully applied to obtain model--independent predictions for the scalar spectral index \cite{Mukhanov:2013tua} as well as for the running \cite{Garcia-Bellido:2014gna} in the canonical single-field inflation scenario.  Interesting results for the excursion $\Delta \phi$ of the inflaton have obtained in \cite{Garcia-Bellido:2014eva} within this formalism.

In practice, the large-$N$ formalism is employed to obtain universal classes of inflationary models from the mathematical relations that hold between two general physical conditions, namely a large number of $e$-folds and the smallness of the slow--roll parameters. In the large-$N$ formalism these two requirements are linked in a single prescription to guarantee a long period of inflation. Concretely, these two conditions are linked through an explicit function of the \emph{Hubble flow slow--roll} parameter $\epsilon_{1}$. An explicit form of  $\epsilon_{1}(N)$ allows one to group families of potentials in a single prescription with common functional forms for the observables. The large-$N$ formalism represents a powerful method of extracting important information about complete classes of inflationary models in a condensed way; as opposed to the usual treatment of individual models starting from an explicit potential.

In this paper we describe the large-$N$ formalism in the context of Tachyon Inflation (Sec.~\ref{Sec:II}). In Sec.~\ref{unisec} we consider three different prescriptions of the $\epsilon_1(N)$ function which cover most of the models considered so far in tachyon inflation. We also find new classes of potentials, to our knowledge not previously published. We also address in two important questions regarding this model: First a check for stability of the model keeping control over the propagation speed for perturbations $c_s$ and the validity of the fluid description by computing the size of entropy perturbations (App.~\ref{app:A}). The second important question is the non-Gaussianity of the different realizations of the model, which we evaluate in Sec.~\ref{unisec} following previous works \cite{Nozari:2013mba}. The large-$N$ formalism also works to directly derive the sets of observables in the tachyon. We contrast the cosmological parameters with the recent observations by  the joint PLANCK-KECK-BICEP analysis \cite{Ade:2015tva}, and the BAO-PLANCK data \cite{BAO}, at second order in the slow--roll parameters (see Sec.~\ref{Sec:Observations}). We use the derived values to demonstrate the relevance of our work for a specific model derived from string theory (Sec.~\ref{subsec:example}). Finally, Section~\ref{Sec:discuss} contains a discussion of our results and concluding remarks. 
%%%%%%%%%%%%%%%%%%%%%%%
\section{Tachyon Inflation and the large-$N$ Formalism}
\label{Sec:II}
%The tachyon action is usually written in the form
%\begin{equation}
%S_T = - \int d^4 x \sqrt{-g}  V(T)  \left( 1 +
%g^{\mu \nu} \partial_{\mu} T
%\partial_{\nu} T  \right)^{1/2} ,\label{eq:action1}
%\end{equation}
%\noindent where the tachyon $T$ is a real scalar field with dimensions of length with potential $V(T)$.
The scenario of inflation driven by a tachyon field is described by the action
\begin{equation}
%S = \frac{1}{16 \pi G}\int d^4 x\sqrt{-g} \, R - \int d^4 x \sqrt{-g}  V(T)  \left( 1 +
%g^{\mu \nu} \partial_{\mu} T
%\partial_{\nu} T  \right)^{1/2} \,.
S = \int d^4 x\sqrt{-g} \left[ \frac{1}{16 \pi} R -   V(T)  \left( 1 +
g^{\mu \nu} \partial_{\mu} T
\partial_{\nu} T  \right)^{1/2} \right] \,.
\label{eq:action}
\end{equation}
%Where the energy-momentum tensor for a tachyonic field is derived from variations of the matter sector of the action with respect to the metric,
%\begin{equation}
%\label{e-m:tensor}
%T_{\mu\nu}= - g_{\mu\nu} \, V(T) \sqrt{1 +
% (\nabla T)^2} + \frac {V(T)}{\sqrt{1+ (\nabla T)^2}} \,
%\partial_{\mu} T \, \partial_{\nu} T .
%\end{equation}
%Here $(\nabla T)^2$ is short for $g^{\mu\nu}\partial_{\mu} T\partial_{\nu} T$.
Our background spacetime is the usual homogeneous and isotropic Friedmann-Lemaitre-Robertson-Walker (FLRW) with a metric given by $ds^2 = - dt^2 + a(t)^2 d \bf x^2 $. Thus, in what follows a prime denotes a derivative with respect to the scalar field and a dot represents a derivative with respect to cosmic time.  The relevant equations for this scenario are the Klein-Gordon that encodes the dynamics of the field and  the Friedmann equation, given in the background by
\begin{eqnarray}
\ddot T +  3 H \dot T(1-\dot T^2)+({\ln V})'(1-\dot T^2)=0 \,, \label{eq:kg}\\
H^2 = \frac{8\pi}{3} \frac{V}{(1 - \dot T^2)^{1/2}} \,, \label{eq:frie} \\ \nonumber
\end{eqnarray}
%The field equation for the tachyon is the Klein-Gordon equation,
%\begin{equation}
%\ddot T +  3 H \dot T(1-\dot T^2)+({\ln V})'(1-\dot T^2)=0 . \label{eq:kg}
%\end{equation}
%\noindent The other relevant equation for the system is the Friedmann equation, derived combing the above with one of the $ii$ components of Einstein's equations. Considering a flat background we derive
%\begin{equation}
%H^2 = \frac{1}{3M_{\rm Pl}^2} \frac{V}{(1 - \dot T^2)^{1/2}}. \label{eq:frie}
%\end{equation}
where the units employed are such that $M_{\rm Pl}=(8 \pi )^{-1/2}$. The Hubble parameter introduced here is defined as  $H = d \ln[a] /dt$. It is natural to introduce the definition of the number of $e$-foldings as a parametrization of the amount of expansion in the inflationary period:
\beq
\label{def:efolds}
N = \int_t^{t_{\rm eoi}} H (\tilde{t})\, d\tilde{t} \,,
\eeq
where the subindex $_{\rm eoi}$ denotes the end of inflation.% It is obvious that in a monotonically expanding spacetime, $N$ can be adopted as a time coordinate.

The accelerated expansion is usually characterized by a set of slow--roll parameters which directly control the steepness of the inflationary potential. For our purposes
 we find convenient to introduce the alternative \textit{Hubble flow} slow--roll parameters, as introduced in Ref. \cite{Schwarz:2001vv}:
\begin{eqnarray}
\epsilon_0 & \equiv & H_* / H ,\label{eq:defeps1} \\
\epsilon_{i+1} & \equiv & \frac{d \ln |\epsilon_i|}{dN}, \ \ \ i
\ge 0. \label{eq:defeps2}
\end{eqnarray}
Here $H_*$ is the Hubble parameter at some chosen time, and $\dot \epsilon_i = H \epsilon_i \epsilon_{i+1}$. Inflation is guaranteed if $\epsilon_1 < 1$. The Hubble flow functions $\epsilon_i$ are defined in terms of derivatives of $H$ with respect to the number of $e$-folds $N$, and in particular we can write
\beq
N(T) =   \sqrt{\frac{3}{2}} \int_{T}^{{T}_{\rm eoi}} \frac{H}{\sqrt{\epsilon_{1}}}
d\tilde{T} \,. \label{eq:defN}
\eeq
This integral is useful in the application of the large-$N$ formalism.  To first order in slow--roll, these parameters can be related to derivatives of the potential as it is shown in  \cite{tachyon1}, thus a relation between the usual slow roll parameters and the Hubble flow slow--roll parameter is easily obtained  if we consider that, within a slow--roll regime, $3H^2\approx V(\phi)$, see reference \cite{Sasaki:1995aw} for  a detailed  explanation.

%\begin{eqnarray} \epsilon_1   &\simeq&  \frac{M_{pl}^{2}}{2} \frac{V'^{2}}{V^{3}}, \label{eq:eps} \\\epsilon_2   &\simeq&M_{pl}^{2} \left(-2\frac{V''}{V^{2}}+3\frac{V'^{2}}{V^{3}}\right),\\\epsilon_2 \epsilon_3 &\simeq& M_{pl}^{4} \left(2\frac{V'''V'}{V^{4}}-10\frac{V''V'^{2}}{V^{5}}+9\frac{V'^{4}}{V^{6}}\right).\end{eqnarray}

%\noindent The relation to the usual parameters for a canonical scalar field $\phi$, defined in terms of the potential $V(\phi)$, is straightforward if we consider that, within a slow--roll regime, $3H^2\approx V(\phi)$, with the results \cite{Sasaki:1995aw}
%\begin{eqnarray}
%\epsilon _{0}&\simeq& M_{pl}\sqrt{\frac{V_{*}} {V}}, \\
%\epsilon _{1}&\simeq& \epsilon, \\
%\epsilon_{2} &\simeq&-2 \eta-4\epsilon.
%\end{eqnarray}
%\noindent The $N$--formalism consists in expressing the evolution of all cosmological observables in terms of $N$ instead of the field $T$ or time.
In order to solve the system of the Klein-Gordon equation \eqref{eq:kg} plus the Friedmann equation \eqref{eq:frie}, one must provide an explicit form of the potential $V(T)$. An alternative set of evolution equations to the Fiedmann/Klein-Gordon set is the following Hamilton-Jacobi system
\begin{eqnarray}
\label{H-J:1}
H'^{2}&=&\frac{9}{4}H^{4}T-16\pi^{2}V^{2},\\
\label{H-J:2}
H'&=&-\frac{3}{2}H^{2}\dot{T}.
\end{eqnarray}
This and Eq.~\eqref{eq:defeps2} imply that a form for the first slow--roll parameter is $\epsilon_{1}(N)=\frac{3}{2}\dot T^{2}$.

It is thus clear that in this formulation one can specify a function $\epsilon_1(N)$ to fully determine the inflationary observables.
%For example, one can use the time
%transformation $dN=-Hdt$ to show that the slow--roll parameter $\epsilon_{1}(N)$ is related to the Hubble parameter as
%\begin{equation}
%\label{epsilon:H}
%\epsilon_{1}(N)=-\frac{\dot H}{H^2}.
%\end{equation}

The cosmological observables can be written in terms of the $\epsilon_1(N)$ parameter as shown in the equations below
%Integrating Eq.~\eqref{epsilon:H} we see that
%\noindent Also, from the definition of the Hubble flow slow--roll parameters, Eq.~\eqref{eq:defeps2}, we see that
\begin{eqnarray}
\label{eq:H}
  H(N)&=& H_{0} \exp \left \lbrack \int^0_N \epsilon_{1} ( N ) dN \right \rbrack \,, \\
\epsilon_{2}&=&-\frac{1}{\epsilon_{1}}\frac{d\epsilon_1}{dN} \,,\\
\epsilon_{3}\epsilon_{2}&=&\frac{1}{\epsilon_{1}}\frac{d^2\epsilon_1}{dN^2}-\left(\frac{1}{\epsilon_{1}}\frac{d\epsilon_1}{dN}\right)^2 \,.
\end{eqnarray}
%
%\noindent Thus the correspondence between the matter field parametrization and the $N$--formalism is given by
%\begin{equation}
%dT= \sqrt{\frac{2}{3}}\frac{\sqrt{\epsilon_{1}}}{H} dN.
%\label{connection}
%\end{equation}
This last equation shows that the explicit function $\epsilon_1(N)$ and the integration of the Hubble parameter in Eq.~\eqref{eq:H} completely determine the evolution of the tachyon field.

A reparametrization of observables may seem unnecessary at first sight. However, there are two main advantages of determining the evolution of fields via $\epsilon_1(N)$: Firstly, a more general description of the dynamics is obtained without appealing to a particular form of the potential. This leads to classes of the tachyon model grouped in a single description. A second feature is the more accurate description
achieved by employing the Hubble flow slow--roll parameters in terms of $\epsilon_1(N)$. This is achieved because, as opposed to the standard slow--roll treatment,  in the present formalism the scalar kinetic term is encoded within the Hubble flow parameters.

To completely define the inflationary models at hand, it is desirable to specify the explicit form of the potential $V(T)$. The reconstruction for the dependence $V(N)$ is made after imposing
the slow--roll approximation. Then $H^2\simeq V$ implies the form for our potential in concordance with  Eq.~(\ref{eq:H}), that is,
\begin{equation}
\label{eq:VV}
V=V_{0}\exp \left \lbrack \int 2 \epsilon_{1} (N)dN \right \rbrack.
\end{equation}
To establish the dependence $V(T)$ we use \eqref{eq:defN} and the chain rule. 
%complement this last result with a function $N(T)$. First provide a prescription for $\epsilon_1(N)$, then use the last expression above (an equality at first order in slow--roll) to integrate %Eq.~(\ref{connection}). This yields an expression for $N(T)$ as an inverse of
%\begin{equation}
%$T=\int \frac{1}{V(N)} \left[\frac{1}{8\pi}\frac{dV(N)}{dN}\right]^{\frac{1}{2}} dN$.
%\label{eq:phN}
%\end{equation}
%derived directly from 

Finally, we can express observables as functions of the number of $e$-folds $N$, as calculated in \cite{tachyon1}. The Hubble slow--roll parameters are,
\begin{eqnarray}
r&=&16\epsilon_{1}+16\epsilon_{1}\left(C\epsilon_{2}-\frac{2}{3}\epsilon_{1}\right) + \mathcal{O}(\epsilon_i^3), \label{r2}\\
n_{s}&=&1-2\epsilon_{1}-\epsilon_{2}-\left\lbrack2\epsilon_{1}^{2}+\left(2C+\frac{4}{3}\right) \epsilon_{1}\epsilon_{2} +C\epsilon_{2}\epsilon_{3}\right\rbrack + \mathcal{O}(\epsilon_i^3), \label{ns2}\\
n_{t}&=&-2\epsilon_{1}-2\epsilon_{1} \left\lbrack \epsilon_{1}+(1+C)\epsilon_{2}
\right\rbrack + \mathcal{O}(\epsilon_i^3). \label{nt2}
\end{eqnarray}
%\begin{eqnarray}
%r_{1}&=&16\epsilon_{1}, \label{r}\\
%n_{s_1}&=&1-2\epsilon_{1}-\epsilon_{2}, \label{ns}\\
%n_{sk_1}&=&\frac{dn_{s}}{d\ln k}=-2\epsilon_{1}\epsilon_{2}-\epsilon_{2}\epsilon_{3},\\
%n_{t_1}&=&-2\epsilon_{1}\label{nt},\\
%n_{tk_1}&=&\frac{dn_{t}}{d\ln k}=-2\epsilon_{1}\epsilon_{2}, \label{ntk1}
%\end{eqnarray}
where $n_{s}$ is the scalar spectral index, $n_{sk}$ its running and $n_{t}$ the tensor spectral index.
All these quantities are evaluated at the scale at which the perturbations are produced, some $50-60$ $e$-folds before the end of inflation.
Here the constant $C=-2+\ln 2+\gamma\simeq -0.72$ and $\gamma$ is the Euler constant.

 A final but important observable of tachyon inflation is the non-Gaussianity of the model.
This has been an important aspect of non-canonical models of inflation as it is pointed out in Refs.~\cite{Creminelli:2005hu,Komatsu:2010hc}. According to the latest observations \cite{Ade:2015ava},  a model of single field inflation can be ruled out if the primordial non-Gaussianity parameter exceeds $\vert f_{\rm NL}- 0.8\vert > 5.0$ in the local limit configuration.
The prescription for adiabatic perturbations within the tachyon inflationary models is \cite{Nozari:2013mba}
\beq
f_{\rm NL} = -\frac23 \frac{\dot T}{H} \left[\frac{1}{H} \ln\left(\frac{k_f}{H}\right) \left(\frac{2 V'''}{3HV}-\frac{2V'V''}{HV^2} + \frac{4V'^3}{3 H V^3}\right) \right],
\eeq
where $k_F^2 = \frac{3H^2V^2- V''V +V'^2}{V^2},$ is the \emph{freeze-out} momentum scale, which indicates a regime of non-linear self-interaction of the scalar field. We shall use this expression to evaluate the non-Gaussianity in the cases of study. As in that reference, we assume adiabatic perturbations, which dominate the power spectrum as shown in the appendix~\ref{app:A}.

Aiming to observationally distinguish our model from the canonical single field inflation, we compute the well known consistency relation, which links the tensor-to-scalar ratio, the scalar spectral index $n_{s}$, and the tensor spectral index $n_{t}$ \cite[see ][]{Lidsey:1995np, tachyon1, Schwarz:2001vv}. These relations could in principle be tested when observables
are determined with enough accuracy. The difference between the canonical scalar field inflation (CSFI) and the tachyon field inflation (TSFI) can be appreciated only at quadratic
order in slow--roll as mentioned in Ref. \cite{tachyon1}. %
That is,
\begin{eqnarray}
\label{CSFI:consistency}
  \mathrm{CSFI}&&\mathrm{consistency\; relation}\qquad \longrightarrow  \qquad  n_{t}=-\frac{r}{8}\lbrack1-\frac{r}{16}+(1-n_{s})\rbrack + \mathcal{O}(\epsilon_i^3), \qquad\qquad\qquad\qquad
\\\nonumber  &   &  \\
\label{TSFI:consistency}
\mathrm{TSFI} &&\mathrm{consistency\; relation} \qquad\longrightarrow   \qquad   n_{t}=-\frac{r}{8}\lbrack1-\frac{r}{24}+(1-n_{s})\rbrack + \mathcal{O}(\epsilon_i^3).\\\nonumber
%\end{array}
\end{eqnarray}
In the first part of the following Section \ref{unisec}, we compare both the CSFI and TSFI scenarios within the perturbative class of models (where the slow--roll function takes the form $\epsilon_1\sim\frac{1}{N}$).

%%%%%%%%%%
\section{Universality Classes for Tachyon Inflation}
\label{unisec}
In this section we present a classification of the potentials for the tachyon field according to the function $\epsilon_1(N)$. Aside from a smooth function, this parameter must vanish asymptotically for large $N$. With the aid of the large-$N$ formalism it is possible to characterize, for large values of $N$,  different models at once in terms of a single form of the function $\epsilon_1$. Here we propose three classes of models, which stem from three different prescriptions for the $\epsilon_1(N)$ function: a perturbative of the form $\epsilon\simeq \frac{1}{N}$ (Sec.~\ref{Sec:IIIa}), an inverted  polynomial function (Sec.~\ref{Sec:IIIb}) and finally an exponential correspondence (Sec.~\ref{Sec:IIIa}). For each class of models we obtain a explicit expressions for the potential and test the cosmological observable parameters up to second order looking at constrains from the observables of tachyonic inflation. For all these classes we are able to obtain reasonable values and, in all of them, a red-tilted spectrum in accordance with observations and a small amplitude for the non-Gaussianity parameter.
%%%%%%%%%%%%%%%%%%%%%%%%%%%%%%%
\begin{figure*}[h!]
\begin{center}
\subfigure[  ]{\label{fig_1}
\includegraphics[width=5cm]{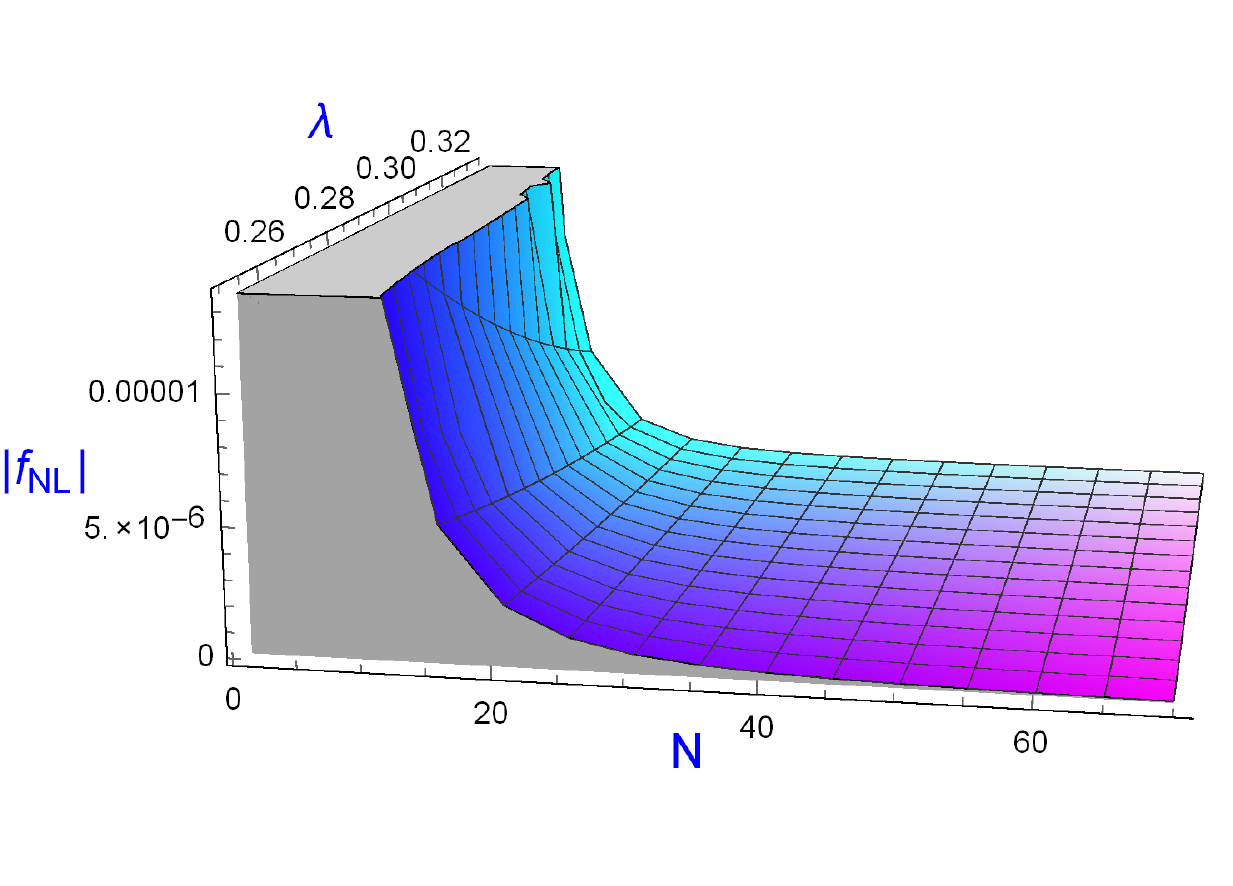}}
\subfigure[  ]{\label{fig_2}
\includegraphics[width=5cm]{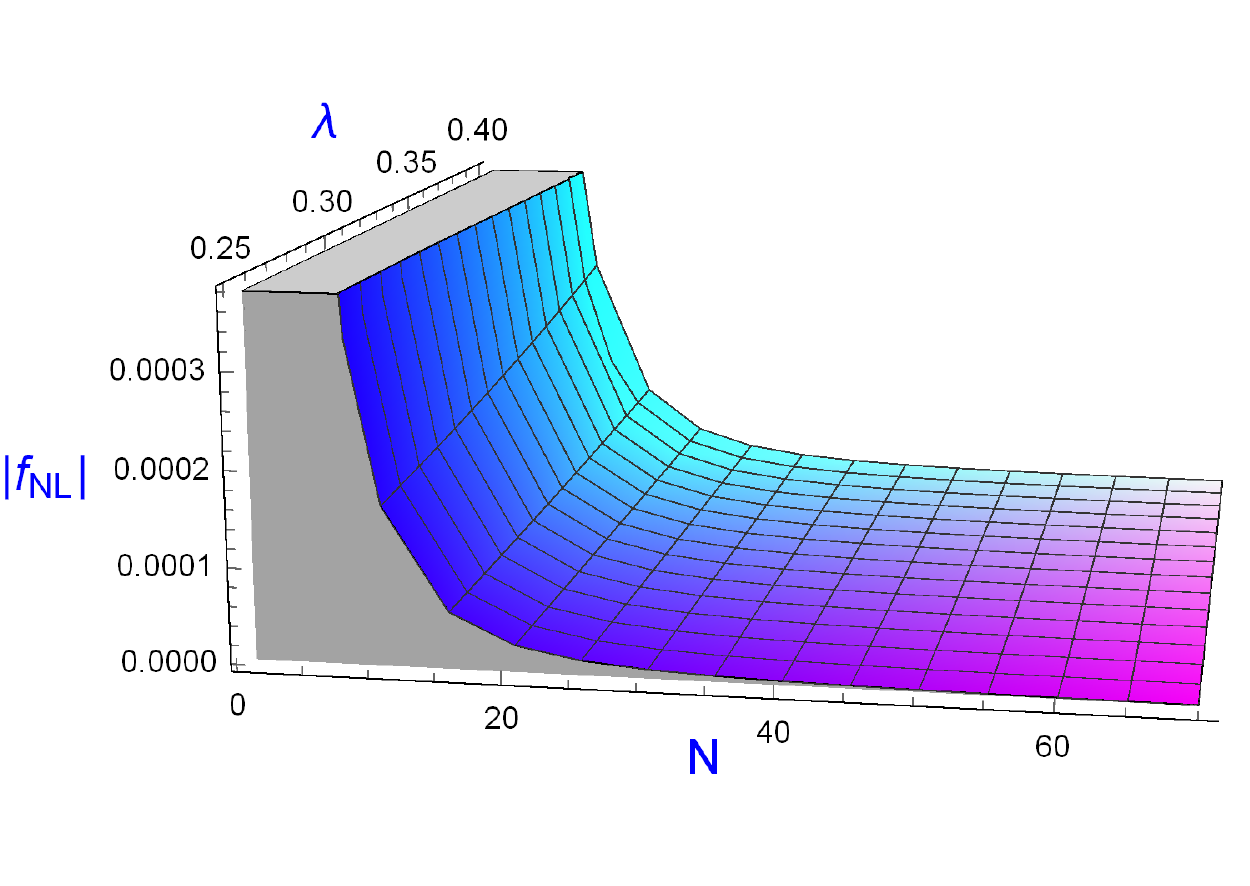}}
\subfigure[]{\label{fig_3}
\includegraphics[width=5cm]{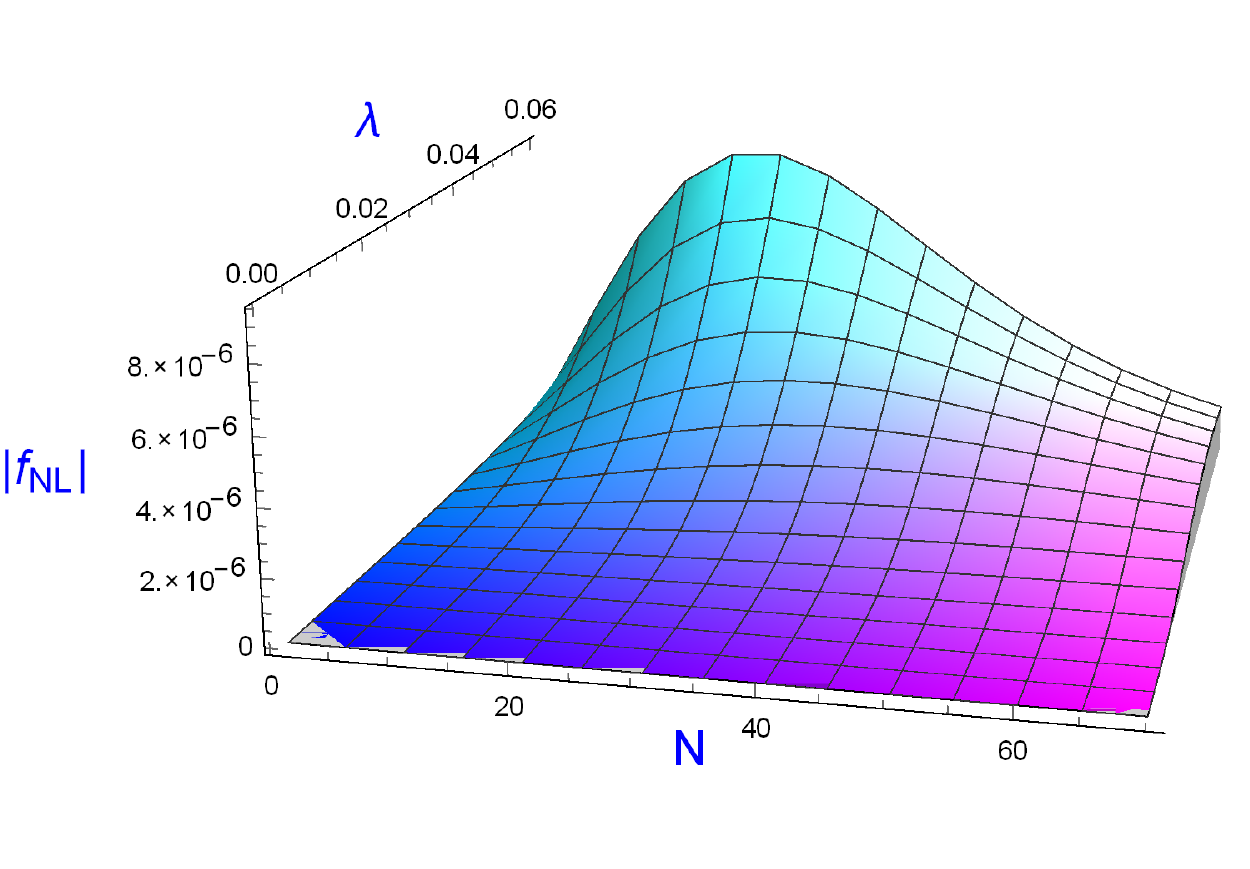}}
\subfigure[]{\label{fig_4}
\includegraphics[width=5cm]{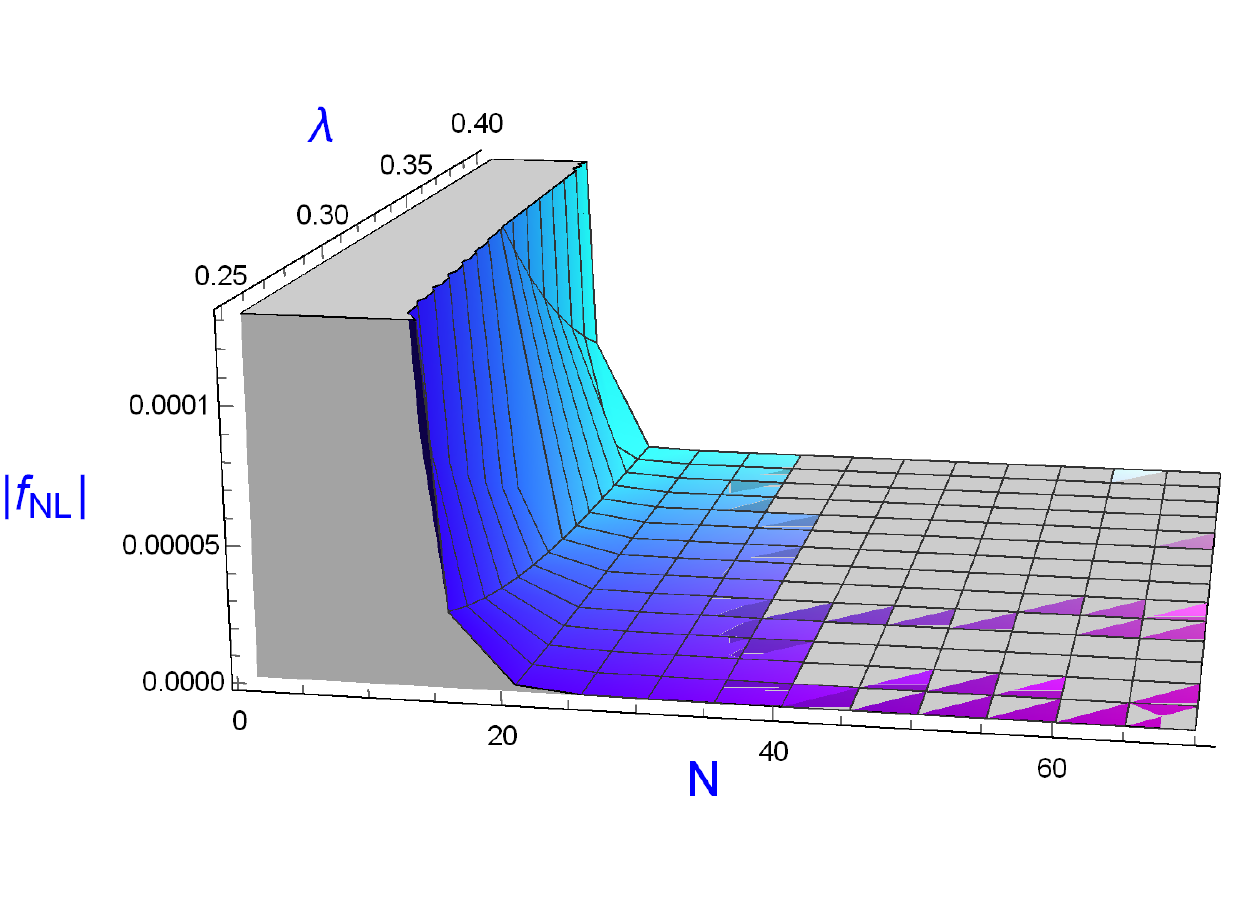}}
\end{center}\vskip -5mm
\caption{We show the behavior of the $f_{\rm NL}$ parameter for the perturbative, polynomial and exponential clases respectively, all of them showing a small value for the non-Gaussianity parameter}
 \label{fig_fnl}
\end{figure*}
%%%%%%%%%%%%%%%%%%%%%%%%%%%%%%%%

In all cases we calculate the value for the speed of sound, as a propagating speed for the scalar perturbations as discussed in Appendix~\ref{app:A}. The condition $c_{s}^2>0$ is fulfilled in all cases, showing that the model is free of pathologies.
%%%%%%%%%%%%%%%%%%%%%%%%%%%%%%%%%%%%%%%%%%%%%%%%%%%%%%%%%%%%%%%%%%%%%%%%%%%%%%%

\subsection{Perturbative class }
\label{Sec:IIIa}
The most direct way to write a small parameter $\epsilon_1$ as a function of the large number of $e$-folds $N$, is the inverse relation, also called the perturbative class, is
\begin{equation}
\epsilon_1 =\frac{\lambda}{N}.
\label{pertclass}
\end{equation}
The family of inflationary potentials in this class is recovered from integrating Eq.~\eqref{eq:VV}; that is, $ V=V_{0}N^{2 \lambda}$. The potential with the usual $T$ dependence and parametrized depending on the value of $\lambda$ reads:
%\begin{equation}
%V= V_{0}\left\lbrack \frac{\pi\left(1-2\lambda\right)^{2}}{\lambda}\right\rbrack^{{\frac{2 \lambda}{ \left(1-2\lambda\right)}}}  T^{\frac{4 \lambda}{1-2 \lambda}}.
%\end{equation}
%\noindent From this general form we find three cases depending on the value of $\lambda$:
\begin{equation}
V(T)=\widehat{V_{0}}T^{\frac{4 \lambda}{1-2 \lambda}} =\begin{cases}
\quad \widehat{V_{0}}T^{n}&\quad0<\lambda<\frac{1}{2},\\ \\
\quad V_{0} \exp\left(\sqrt{8\pi V_{0}}T\right)&\quad\lambda=\frac{1}{2},\\ \\
\quad\widehat{V_{0}}T^{-n}&\quad \frac{1}{2} < \lambda < \frac{3 N}{2},
\end{cases}
\label {cases}
\end{equation}
where $\widehat{V_{0}}=V_{0}\left\lbrack \frac{\pi\left(1-2\lambda\right)^{2}}{\lambda}\right\rbrack^{{\frac{2 \lambda}{ \left(1-2\lambda\right)}}}$ and $n=\mid\frac{4\lambda}{1-2\lambda}\mid$.
We immediately see that from this perturbative class the monomial power law potential is recovered for $\lambda<\frac{1}{2}$. In particular, the quadratic and quartic forms,
widely analyzed in the literature (e.g.~\cite{Nozari:2013mba,tachyon2} and references therein), are recovered in this class of the large-$N$ formalism.
    The upper bound $\lambda = \frac{3 }{2} N$ in the third family of potentials is set to guarantee that the effective sound speed $c_s$ be well defined. Indeed, for this class of models is given by
\begin{equation}
c_{s}^2={1-\frac{2 \lambda}{3 N}},
\end{equation}
which limits the range of values of $\lambda$ in Eq.~(\ref{cases}).

%\begin{equation}
%n_{s\,{1}}=1-\frac{1+2\lambda}{N},\qquad\qquad r_{1}=\frac{16
%\lambda}{N},\qquad\qquad
%n_{sk\,{1}}=-\frac{1}{N^2}-\frac{2\lambda}{N^2},\qquad\qquad
%n_{tk\,{1}}=-\frac{2\lambda}{N^2} .
%\end{equation}

%%%%%%%%%%%%%%%%%%%%%%%%%%%%%%%%%%%%%%%%%%%%%%%%%%%%%%%%%%%%%%%%%%%%%
\begin{table}[tbp]
  \centering
\begin{tabular}{| c| c| c | c | c | c | c| }
  \hline
 \small { $V(T)\sim$} & $\lambda$ & $n_{s}$ & $n_{t}$& $r$  & $c_{s}$ & $ \mid{f_{\rm NL}}\mid$\\
  \hline
   $T^{2}$& $0.25$& $0.9699$ & $-0.0101$ & $0.0787$ & $0.9983$&$1.6\times 10^{-7}$ \\
  \hline
   $T^{4}$& $0.3333$ & $0.9665$& $-0.0134$& $0.1048$ & $0.9977$&$4.9\times 10^{-8}$ \\
  \hline
   $T^{6}$& $0.375$& $0.9648$& $-0.0151$& $0.1179$ & $0.9974$ &$2\times 10^{-8}$\\
  \hline
   $T^{8}$& $0.4$& $0.9637$ & $-0.0162$& $0.1258$ & $0.9973$& $1\times 10^{-8}$\\
  \hline
  \tiny{$ \exp\left(\sqrt{8\pi V_{0}}T\right)$}& $.05$& $0.9595$ & $-0.0203$& $0.1571$ & $0.9966$ &$0$\\
  \hline
\end{tabular}
\bigskip
\caption{Values of the observables are here given for potentials determined by the parameter $\lambda$ in the perturbative class,
up to second order in the Hubble flow parameters and at $N=50$ $e$-folds of inflation.}
\label{tableper2}%
\end{table}%
%%%%%%%%%%%%%%%%%%%%%%%%%%%%%%%%%%%%%%%%%%%%%%%%%%%%%%%%%%%%%%%%%%%%%%%%%%

The spectral index, the tensor-to-scalar ratio, as well as the scalar and tensor runnings to second order in slow--roll can be derived from Eqs.~\eqref{r2}~to~\eqref{nt2}. Explicitly,
\begin{equation}
r=\frac{16 \lambda}{N} +\frac{\left(C-\frac{1}{3}\lambda \right)}{N^{2}},\qquad
n_{s}=1-\frac{1+2\lambda}{N}-\frac{2\lambda(\lambda+2C+\frac{8}{3})}{N^{2}},\qquad
n_{t}=-\frac{2\lambda}{N}-\frac{2\lambda(\lambda+2C+1)}{N^2}.
\end{equation}

When evaluating the observables at $N=50$ we obtain the
cosmological parameters values given in Table~\ref{tableper2}. Note that for the perturbative class we obtain values for $r$ small enough to meet the Bicep2/Keck/Planck constraints. The numerical values of cosmological parameters will appreciably change when varying $N$, an shown in Section~\ref{Sec:Observations}.

In a rough comparison between the CSFI and TSFI scenarios within the perturbative class, we note that even at first order there are differences between the numerical values of the cosmological parameters between both models (see Table~\ref{tablecomp}). On the other hand, the difference between the consistency relations of Eqs.~\eqref{CSFI:consistency} and \eqref{TSFI:consistency} shows that
\begin{equation}
\Delta n_{t}\equiv n^{CSFI}_{t}-n^{TSFI}_{t}=-\frac{r}{48}
\label{DCT}
\end{equation}

If future experiments measure the tensor-to-scalar ratio and the tensor index, a precision of order $1/50$  is required in the latter with respect to the face value of $r$ in order to distinguish between the CSFI and TSFI scenarios. Regarding the differences kept in other observables, we show in the next section that TSFI provides a better fit to the current data than the CSFI.

%%%%%%%%%%%%%%%%%%%%%%%%%%%%%%%%%%%%%%%%%%%%%%%%%%%%%%%%%%%%%%%%%%%%%%%%%%
\begin{table}[tbp]
  \centering
\begin{tabular}{| c| c| c | c | c |c |c |}
  \hline
 \small { $V(T)$ \text{vs} $V(\phi)$}  & $n_{s\,{1}}(T)$ & $n_{s\,{1}}(\phi)$& $\mid{n_{s\,{1}}(T)-n_{s\,{1}}(\phi)}\mid $ & $r_{1}(T)$ & $r_{1}(\phi)$ & $\mid{r_{1}(T)-r_{1}(\phi)}\mid $\\
  \hline
   $T^{2}~~ \text{vs}~~  \phi^{2}$& $0.97$& $0.96$ & $0.01 $& $0.08 $ & $0.16 $& $0.08$\\
  \hline
   $T^{4}~~ \text{vs}~~ \phi^{4}$& $0.9666$ & $0.94$& $0.02668 $& $0.10656$ & $0.32 $& $0.21344$\\
  \hline
   $T^{6}~~ \text{vs}~~ \phi^{6}$& $0.965$& $0.92$& $0.045 $& $0.12 $& $0.48 $& $0.36 $\\
  \hline
   $T^{8}~~ \text{vs}~~ \phi^{8}$& $0.964$& $0.9$& $0.064 $& $0.128 $& $ 0.64$& $ 0.512$\\
  \hline
\end{tabular}
\bigskip
\caption{Comparison between the canonical single field (CSFI) and the tachyonic (TSFI) inflationary scenarios. Cosmological parameters at first order are evaluated for $N=50$ $e$-folds of inflation.}
\label{tablecomp}%
\end{table}%
%%%%%%%%%%%%%%%%%%%%%%%%%%%%%%%%%%%%%%%%%%%%%%%%%%%%%%%%%%%%%%%%%%%%%%%%%%5
%%
%\begin{figure*}[htb]
%\begin{center}
%\includegraphics[width=10cm]{pertcase}
%\end{center}\vskip -5mm
%\caption{We show the behavior of the $f_{\rm NL}$ parameter for the perturbative class with equation of state parameter $\epsilon_1=\frac{\lambda}{N}$.}
 %\label{perturbative}
%\end{figure*}
%%
The non-Gaussianity parameter for the perturbative class of models is given in the local limit by
\begin{equation}
f_{\rm NL}=\frac{\left(1-2\lambda\right)^2}{3N^2} \ln  \left(\frac{1-2\lambda+2N}{2N}\right).
\end{equation}
The amplitude of non-Gaussianity is plotted in Figure~\ref{fig_1}. As the equation above shows, $f_{\rm NL}$ is small since it behaves like $\frac{1}{N^2}$, regardless of the value of $\lambda$. Thus, for large-$N$, $f_{\rm NL}$ will always present negligible values.

%%%%%%%%%%%%%%%%%%%%%%%%%%%%%%%%%%%%%%%%%%%%%%%%%%%%%%%%%%%%%%%%%%%%%%%%%%%%%%%%%%%%%%%%%%%%%%%%%%%%

\subsection{Polynomial class}
\label{Sec:IIIb}
The polynomial class is characterized by an equation of state parameter given by
\begin{equation}
\epsilon_1=\frac{\lambda}{N( N^{2\lambda}+1)}.
\end{equation}
While previous works have dismissed this class of models, arguing that sub-leading terms in the denominator provide a negligible contribution, here we show sensible differences can be distinguished between this and the previous perturbative class.

The family of inflationary potentials in this class is takes the form $ V=V_{0}\frac{N^{2 \lambda}}{N^{2 \lambda}+1}$. Consequently, the potential as a function of $T$ is shown in the equation below depending on the value of $\lambda$
%\begin{equation}
%V= V_{0} \left \lbrack1-\frac{1}{V_{0}\left( \frac{\pi V_{0}\left(1-2\lambda\right)^{2}}{\lambda}\right)^{{\frac{2 \lambda}{ \left(1-2\lambda\right)}}}  T^{\frac{4 \lambda}{1-2 \lambda}}+1} \right \rbrack.
%\end{equation}
%From this form of the potential we obtain three cases depending on the value of $\lambda$
\begin{equation}
\label{polynomial:potentials}
%\small{
V(T)=  \left \lbrack V_{0}-\frac{1}{\left( \frac{\pi V_{0}\left(1-2\lambda\right)^{2}}{\lambda}\right)^{{\frac{2 \lambda}{ \left(1-2\lambda\right)}}}  T^{\frac{4 \lambda}{1-2 \lambda}}+1} \right \rbrack = \begin{cases}
V_{0}\frac{AT^{n}}{AT^{n}+1}&\!\!\!\!\!\!\!\!\!\!\!\!\quad0<\lambda<\frac{1}{2},\\ \\
\small{\frac{1}{2}V_{0} \left\lbrack 1+\tanh \left(\sqrt{2 \pi V_{0}}T\right) \right\rbrack}&\!\quad\lambda=\frac{1}{2},\\ \\
V_{0}\frac{A}{A+T^{n}}&\!\quad\lambda>\frac{1}{2},
\end{cases}
%}
\end{equation}
where $A=\left\lbrack \frac{V_{0}\pi\left(1-2\lambda\right)^{2}}{\lambda}\right\rbrack^{{\frac{2 \lambda}{ \left(1-2\lambda\right)}}}$ and $n=\mid\frac{4\lambda}{1-2\lambda}\mid$. {To our knowledge, this class of potentials has not been explored in previous studies of Tachyonic Inflation. }
%The observables at first order are given by
%\begin{eqnarray}
%\nonumber n_{s\,{1}}&=&\frac{-1-2\lambda+N}{N},\\ \nonumber
%_{1}&=&\frac{16 \lambda}{N(N^{2\lambda}+1)},\\ \nonumber
%n_{sk\,{1}}&=&-\frac{1}{N^2}-\frac{2\lambda}{N^2}, \\
%n_{tk\,{1}}&=&-\frac{2\lambda\left\lbrack1+N^{2\lambda}(1+2\lambda)\right\rbrack}{\left\lbrack
%N(N^{1+2\lambda}+1)\right\rbrack^2},
%\end{eqnarray}
While the effective sound speed  $c^{2}_{s}=1-\frac{2}{3}\epsilon_{1}$
is a real number regardless of values for $\lambda$ in a realistic model.
%%
%\begin{figure}
%\begin{minipage}{.49\linewidth}
%\centering
%\includegraphics[width=10cm]{polycase}
%\end{minipage}
%\caption{The equation of state parameter specified by
% $\epsilon_1=\frac{\lambda}{N(N^{2\lambda}+1)}$ for the polynomial class yields a very small value of the $f_{\rm NL}$ parameter
   %in the large-$N$ limit.}
%\label{polyclass}
%\end{figure}
%%
Computed up to second order in the Hubble flow parameters (up to $\mathcal{O}(\epsilon_i^3)$),  the values for the spectral and tensor indices and the tensor to scalar ratio are
\begin{eqnarray}
\nonumber
%n_{s}&=&-\left\lbrack\frac{3(C+N-N^2)(1+N^{2\lambda})^2+2\lambda^2\left\lbrack
%3+2N^{2\lambda}\left(8+3C-2\right)\right\rbrack+2\lambda\left(1+N^{2\lambda}\right)\left\lbrack
%8+6C+3N+3N^{2\lambda}(C+N)\right\rbrack}{N^2\left(1+N^{2\lambda}\right)^2}\right\rbrack,\\ \nonumber
n_{s}&=&-\left\lbrack\frac{3(C+N-N^2)(1+N^{2\lambda})^2+2\lambda^2\left\lbrack
3+2N^{2\lambda}\left(8+3C-2\right)\right\rbrack}{N^2\left(1+N^{2\lambda}\right)^2}\right\rbrack
\\ \nonumber
&-&\left\lbrack\frac{2\lambda\left(1+N^{2\lambda}\right)\left\lbrack
8+6C+3N+3N^{2\lambda}(C+N)\right\rbrack}{N^2\left(1+N^{2\lambda}\right)^2}\right\rbrack,
\\ \nonumber
n_{t}&=&\frac{-2\lambda\left(1+\lambda+C+N\right)-2\lambda N^{2\lambda}\left\lbrack1+C+2\lambda(1+C)+N\right\rbrack}{N^2\left(1+N^{2\lambda}\right)^2},\\
r&=&\frac{16\lambda\left\lbrack -\lambda+3(N+C)+3N^{2\lambda}\left(C+2\lambda C+N\right)\right\rbrack}{3N^2\left(1+N^{2\lambda}\right)^2}.
\end{eqnarray}
%%%%%%%%%%%%%%
\begin{table}[tbp]
  \centering
\begin{tabular}{| c | c | c | c | c | c | c | }
\hline
 \small {$V(T)\sim$}& $\lambda$ & $n_{s}$& $n_{t}$ & $r$  & $c_{s}$ &$\mid{f_{\rm NL}}\mid$\\
  \hline
 \tiny{ $\frac{A T^{2}}{A T^{2}+1}$}& $0.25$  & $0.9703$ & $-0.0012$& $0.0097$&$0.9997$&$2.2\times 10^{-6}$\\
  \hline
   \tiny{ $\frac{A T^{4}}{A T^{4}+1}$}& $0.3333$ & $0.967$ & $-0.0009$& $0.0071$& $0.9998$&$3\times 10^{-6}$ \\
  \hline
   \tiny{ $\frac{A T^{6}}{A T^{6}+1}$}& $0.375$ & $0.9654$& $-0.0007$& $0.0059$ &$0.9998$ &$3.4\times 10^{-6}$\\
  \hline
   \tiny{ $\frac{A T^{8}}{A T^{8}+1}$}& $0.4$  & $0.9644$ &$-0.0006$& $0.0052$& $0.9998$&$3.7\times 10^{-6}$ \\
  \hline~
   \tiny{ $\!\!\!\! \!\!1 \!\!\!~+ \! \tanh \!\left(\!\sqrt{2 \pi V_{0}}T \right)$}& $0.5$  & $0.9605$ & $-0.0003$& $0.003$&$0.9999$&$4.8\times 10^{-6}$ \\
\hline
\end{tabular}
\bigskip
\caption{Values for observables in the polynomial class at second order in the Hubble flow parameters for the some $\lambda$ values  for $N=50$ $e$-folds of inflation.}
\label{tablepol2}%
\end{table}
%%%%%
Numerical results for observables for $N=50$ are given in Table~\ref{tablepol2}. Finally, the non-Gaussianity parameter for this class is
\begin{equation}
f_{\rm NL}=\left\lbrack
\frac{(1-2\lambda)^2+N^{4}\lambda(1+2\lambda)-2N^{2\lambda}\left(\lambda+6\lambda^2-1\right)}{3N^2\left(1+N^{2\lambda}\right)^2}\right\rbrack \ln  \left( 1-\frac{2\lambda}{N(1+N^{2\lambda})}+\frac{1+2\lambda}{2N}\right).
\end{equation}
As in the case of the perturbative class, Figure~\ref{fig_2} shows that $f_{\rm NL}$ is small and  asymptotically vanishing like  $\frac{\lambda}{N(N^{2\lambda}+1)}$.

%%%%%%%%%%%%%%%%%%%%%%%%%%%%%%%%%%%%%%%%%%%%%%%%%%%%%%%%%%%%%%%%%%%%%%%%%%%%%%%%%%%%%%%%%%%%%%%%%%%%%
\subsection{Exponential class}
\label{Sec:IIIc}
As a final example of the versatility of the large-$N$ formalism, we present two specific cases where the explicit form of the potential in terms of the field can be obtained. From the first
case we recover a potential previously found in Refs.~\cite{Sami:2002fs,Sen:2002an, tachyon1} for a tachyon inflationary model. The results obtained here complement those studies and provide important constraints on their parameters. In the second case we find a potential of the eternal inflation type. In both cases all the observable parameters are computed explicitly.
\subsubsection{First exponential case}
The first class is characterized by an equation of the state parameter given by
\begin{equation}
\epsilon_{1} =\frac{\lambda}{2\left( e^{\lambda N}+1\right)}.
\end{equation}
%and that the first order observables are directly computed as
%\begin{eqnarray}
%\nonumber n_{s\,{1}}&=&1-\lambda,\qquad\qquad r_{1}=\frac{8 \lambda}{
%e^{\lambda N}+1},\\ \nonumber
%n_{sk\,{1}}&=&0, \qquad\qquad\qquad n_{tk\,{1}}=-\frac{1}{4} \lambda^{2}\text{sech}^2 \left(\frac{\lambda N}{2}\right) .\\
%\end{eqnarray}
The family of inflationary potentials in this class is integrated as $  V=V_0 \frac{ e^{\lambda N}}{e^{\lambda N}+1}$, and the explicit form as a function of the tachyon field reads
\begin{equation}
V=\frac{V_0}{2 \pi\lambda V_0 T ^2+1}.
\end{equation}

%%lalala
%\begin{figure}
%\begin{minipage}{.49\linewidth}
%\centering
%\includegraphics[width=10cm]{expcasetwo}
%\end{minipage}
%\caption{ The exponential class characterized by the equation of state $\epsilon =\frac{\lambda}{2\left( e^{\lambda N}+1\right) } $,
%as function of $N$ and $\lambda$.}
%\label{expcase1}
%\end{figure}
%%%

The effective speed of sound and the values for the observables up to order $\mathcal{O}(\epsilon_i^3)$ in the Hubble flow parameters are given by
\begin{eqnarray}
c_{s}^2&=&{1-\frac{\lambda}{3 \left( e^{\lambda N}+1\right)}},\\
\nonumber
n_{s}&=&1-\lambda-\frac{\lambda^{2}\left\lbrack-3-2e^{\lambda
N}(8+3C)\right\rbrack}{6(1+e^{\lambda N})^{2}} + \mathcal{O}(\epsilon_i^3),\\
\nonumber
n_{t}&=&- \frac{\lambda\left\lbrack2+\lambda+2(1+\lambda+\lambda
C)e^{\lambda N} \right\rbrack}{(1+e^{\lambda N})^{2}}+ \mathcal{O}(\epsilon_i^3),\\
 \nonumber
r&=&\frac{4 \lambda \left\lbrack 6-\lambda+6(1+\lambda C)e^{\lambda
N} \right\rbrack}{3\left(1+e^{\lambda N} \right)^{2}} + \mathcal{O}(\epsilon_i^3).
\end{eqnarray}

From plotting the non-Gaussianity parameter,
\begin{equation}
   f_{\rm NL} = \frac{\lambda^2(1-3e^{\lambda N})\text{log}\left( 1+\frac{\lambda}{2} \text{tanh}\left( \frac{\lambda N}{2} \right) \right)}
   {3\left( 1+ e^{\lambda N} \right)^2},
\end{equation}
we note that the $f_{\rm NL}$ shows a peak for $N \sim 30$ and then vanishes asymptotically (Figure~\ref{fig_3}), while the value for the first non-Gaussianity parameter is again small, this might be a feature worth exploring for the cases where $f_{\rm NL}$ could be large.

\subsubsection{Second exponential case}
The second class is characterized by the following equation of the state parameter
\begin{equation}
\epsilon_1 =\frac{2\lambda}{e^{2\lambda N}-e^{-2\lambda N}}.
\end{equation}
%with observables given by
%\begin{eqnarray}
%\nonumber
%n_{s\,{1}}&=&1-2\lambda \coth (\lambda N),\qquad\qquad r_{1}=16 \lambda\, \text{csch}(2 \lambda N),\\
%n_{sk\,{1}}&=&-2 \lambda^{2}\,\text{csch}^{2}(\lambda N),\qquad\qquad
%n_{tk\,{1}}=-4 \lambda^{2} \coth (2 \lambda N)\, \text{csch}(2 \lambda
%N).
%\end{eqnarray}
The family of potentials $  V=V_0 \tanh (\lambda N)$, given in terms of $V(T)$ is
\begin{equation}\label{eq:sech}
V=V_{0}\,\frac{1}{\cosh\left( \sqrt{8 \pi  \lambda V_0}T \right)} .
\end{equation}

This is one of the most popular models in the literature, \cite[e.g.][]{cosh,Daniel}. While our approach derives potentials from expressions of $\epsilon_1$, we show below that our analysis and subsequent fit to observations can contribute to constrain models inspired in string theory \cite{Daniel}. 

%\texttt{El potencial  de inflaci�n eterna surgido de esta clase exponencial
%es uno de los mas populares en la literatura, ver por ejemplo
%\cite{cosh,Daniel}, en nuestro an�lisis lo hemos construido pensando
%en una clase exponencial para $\epsilon_1$, por ello pensamos que es
%interesante hacer una comparaci�n directa con su an�logo surgido de
%la teor�a de cuerdas \cite{Daniel} y ver que pasa cuando los
%ajustamos a las observaciones.}

In this class the effective sound speed and observables up to order $\mathcal{O}(\epsilon_i^3)$ in the Hubble parameter are
\begin{eqnarray}
c_{s}^2&=&{1-\frac{2}{3} \lambda\, \text{csch}(2\lambda N)},\\
\nonumber n_{s}&=&1-\lambda \,\text{csch}^{2}(2 \lambda N)\left\lbrack
\frac{4}{3}\lambda(8+6C) \cosh(2\lambda N)+2(\lambda+2\lambda
C+\sinh(2\lambda N))+\sinh(4\lambda N)\right\rbrack, \\ \nonumber
n_{t}&=&- 2\lambda \,\text{csch}(2 \lambda N)\left\lbrack
1+2\lambda(1+C)\coth(2\lambda N)+\lambda \,\text{csch}(2 \lambda N)
\right\rbrack, \\ \nonumber r&=&16 \lambda\,\text{csch}(2 \lambda
N)\left\lbrack 1+2\lambda\coth(2\lambda N)-\frac{1}{3}\lambda
\,\text{csch}(2 \lambda N)\right\rbrack.
\end{eqnarray}

%%%
%\begin{figure}
%\begin{minipage}{.49\linewidth}
%\centering
%\includegraphics[width=10cm]{expcaseone}
%\end{minipage}
%\caption{ The exponential class characterized by the equation of state $\epsilon_1 =\frac{2\lambda}{e^{2\lambda N}-e^{-2\lambda N}} $  is here illustrated as function of $N$ and $\lambda$.}
%\label{expcase2}
%\end{figure}
%%%
If we take this results and evaluate for $N=50$ we obtain the
cosmological parameters values given in Table \ref{tableexp2}. The
non-Gaussianity for this class is
\begin{equation}
f_{\rm NL}= -\frac{4}{3} \lambda^2\, \text{sech}^2(\lambda N) \ln \left\lbrack1+\lambda \tanh (\lambda N)\right\rbrack,
\end{equation}
with $f_{\rm NL}$ again small as shown in Figure~\ref{fig_4}.
%%%%%%%%
\begin{table}[tbp]
  \centering
\begin{tabular}{| c | c | c | c | c | c | c |  }
\hline
 \small{ $V(T)\sim$}& $\lambda$ & $n_{s}$& $n_{t}$ & $r$  & $c_{s}$ &$\mid{f_{\rm NL}}\mid$ \\
  \hline
   \tiny{ $\!\!\!\,\frac{1}{\cosh\left( \sqrt{8 \pi  \lambda V_0}T \right)}$}& $1\times 10^{-7}$ & $0.9595$  & $-0.0203$ & $0.1571$ & $0.9966$ &$6.6\times 10^{-27}$\\
  \hline
   \tiny{ $\frac{1}{2 \pi\lambda V_0 T ^2+1}$}& $.04$ & $0.9596$&  $-0.0048$   & $0.0371$ & $0.9992$&$2.4\times 10^{-6}$  \\
  \hline
\end{tabular}
\bigskip
\caption{Observables in both cases of the exponential class of models, at second order in the Hubble flow parameters, for $N=50$ $e$-folds of inflation.}
\label{tableexp2}
\end{table}

%%%%%%%%%%%%%%%%%%%%%%%%%%%%%%%%%%%%%%%%%%%%%%%%%%%%%%%%%%%%%%%%%%%%%%%%%%%%%%%%%%%%%%%%%%%%%%%%%%%%%%
\section{Cosmological Parameters Vs Observations}
\label{Sec:Observations}

\subsection{Parameter fitting}
\begin{figure*}[h!]
\begin{center}
\subfigure[ The inflation models correspond to the perturbative class with potentials
  $V\propto T^2$ (blue),
$T^4$ (green), $T^6$ (red), $T^8$ (yellow),
and $e^{\sqrt{8\pi V_0}T}$ (cyan). ]{\label{fig_6}
\includegraphics[width=6cm]{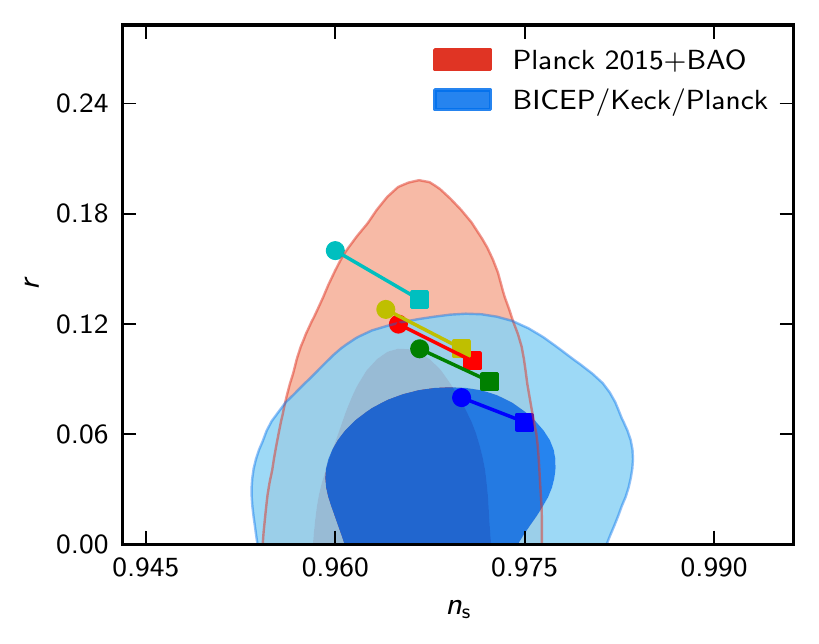}}
~~~~~\subfigure[The inflation
models correspond to the polynomial class with potentials $V\propto~AT^n/(AT^n+1)$ for $n=2$ (blue),
$n=4$ (green), $n=6$ (red), and $n=8$
(yellow).]{\label{fig_7}
\includegraphics[width=6cm]{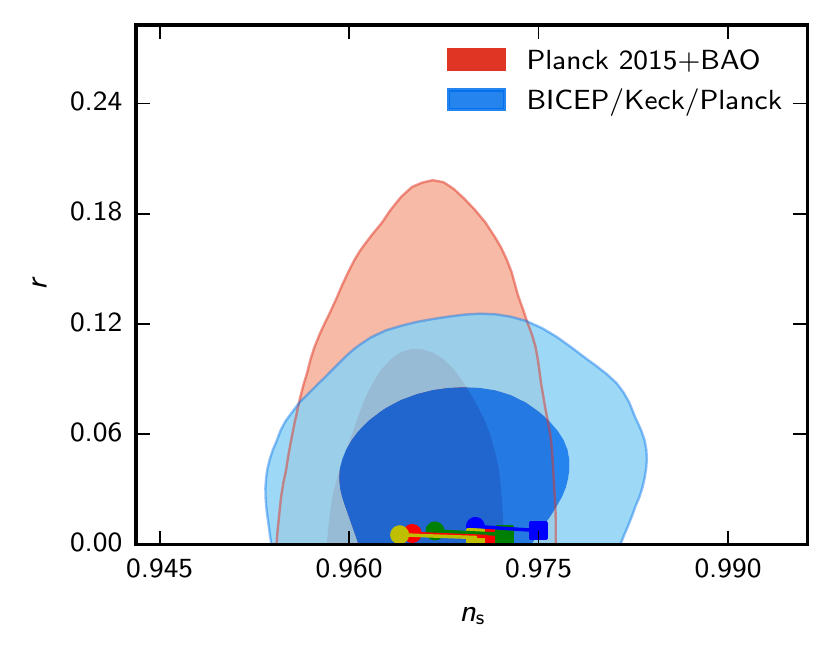}}
\subfigure[Inflation models corresponding to the exponential classes. The potentials are
   $V\propto 1/\cosh(\sqrt{8\pi\lambda}T)$ for $\lambda=0.01$
(blue), and $\lambda=10^{-7}$ (green). Also
$V=V_0/(2\pi \lambda V_0 T^2+1)$ for $\lambda=0.02$ (red), and
$\lambda=0.0307$ (yellow), see Table~\ref{tableexp2}.]{\label{fig_8}
\includegraphics[width=6cm]{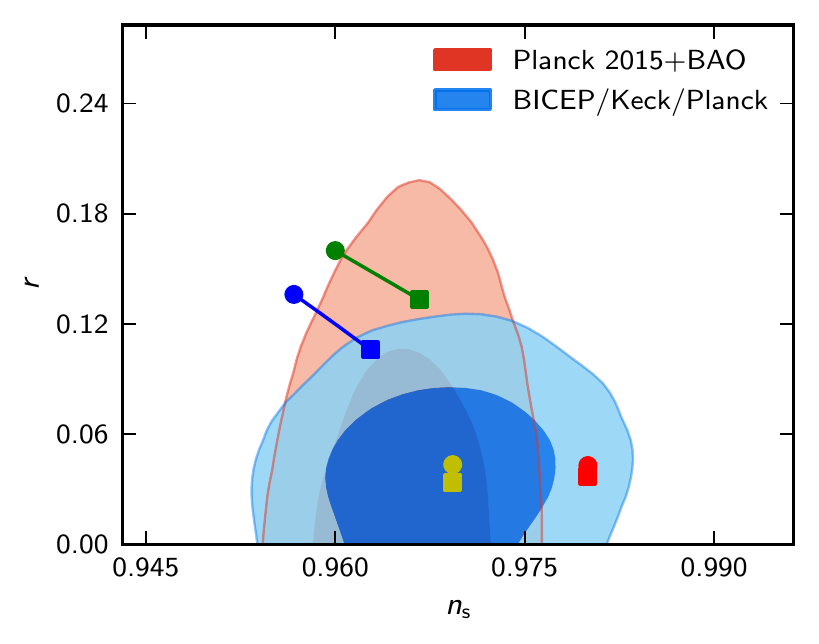}}
\end{center}\vskip -5mm
\caption{Marginalized confidence regions at $68\%$ and $95\%$ for $n_s$ and
$r$ for the different classes of potentials in tachyon inflation. The blue region corresponds to Planck 2015 TT, low P and lensing data \cite{Adam:2015rua} plus the BICEP2/Keck/Planck collaboration
data \cite{Ade:2015tva}. The red region represents Planck
2015 TT, TE, EE, low P and lensing confidence regions \cite{Adam:2015rua} including BAO
from ref. \cite{BAO}. $\blacksquare$ indicates evaluation of models at $N=60$ and thick coloured lines denote intermediate values up to the lower value $N= 50 $, represented by $\bullet$.}
 \label{fig_FermionIII}
\end{figure*}

We have used a suite of cosmological data to compare different models with observations. In Figure~\ref{fig_FermionIII} we present the marginalized confidence regions at 68\% and 95\%
on the pair of parameters $(n_s,\,r)$ from two sets of observations. The red contours correspond to the confidence regions obtained from CMB data from
the Planck satellite \cite{Adam:2015rua} plus
BAO data from different observations at redshifts $z=0.106, 0.15, 0.57$ and $0.32$.
We employed the 2015 Planck data release including TT, TE, EE, low P and lensing data.
%As stated in the Planck papers, beam mismatch generates leakage from temperature to polarization
%power spectrum and the results from TE and EE Planck polarization data
%should be treated with care until leakage corrections are implemented in future
%data releases.
On the other hand, the
blue contours correspond to
baseline and lensing Planck data combined with Bicep2/Keck/Planck joint constraints
on B-mode polarization \cite{Ade:2015tva}.

For both datasets we used the publicly available Markov chains from Planck,
analyzed by using python scripts from CosmoMC \cite{Lewis:2002ah}.
These chains were produced considering $n_s$, $n_{sk}$ and $r$ as free parameters.
%considering a cosmological model with 8 parameters including
%6 parameters from the minimal $\Lambda CDM$ model $\Omega_b h^2$, $\Omega_c h^2$
%$\theta_{mc}$, $\tau$, $A_s$, $n_s$ and the two extra parameters $n_{sk}$ and $r$.
The primordial scalar power spectrum is determined by $A_s$, $n_s$ and  $n_{sk}$, while
the primordial tensor power spectrum will have an amplitude determined by $r$, with
$n_t$ satisfying the consistency relation (\ref{TSFI:consistency}) at first order.
%with $n_t=1$ and $n_{tk}=0$. As can be seen from the tables
%in section \ref{unisec}, the values of those parameters are not zero, and
%a complete, more accurate analysis should take them into account.
%%%%%%%%%%%
%%%%%%%%%%%%%%%%%%%

The results of our parameter analysis are condensed in Figure~\ref{fig_FermionIII}. From the Markov
chains we obtain the primordial power spectrum parameters to have values of $n_s=0.966\pm 0.008$,
$n_{sk}=-0.008\pm0.015$ and $r<0.17$ at 95\% confidence for the Planck 2015 plus BAO data and
$n_s=0.968\pm 0.008$,
$n_{sk}=-0.007\pm0.015$ and $r<0.11$ at 95\% confidence for the
BICEP2/Keck/Planck data. In the leftmost panel,  Figure~\ref{fig_6},
the perturbative class is reported. The model with exponential potential $e^{\sqrt{8\pi \lambda V_{0}}T}$
(corresponding to $\lambda=1/2$)  is
excluded at 95\% confidence for the BICEP2/Keck/Planck data.
In the same panel, the models corresponding to potentials of the form $V(T)\propto T^4$, $T^6$ and $T^8$ fall outside the 68\% confidence region of both data sets. Such tension might increase with future CMB polarization observations.
In the rightmost panel, Figure~\ref{fig_7}, we see that all the
polynomial models studied fall inside the 68\% confidence region of both datasets, at
least for some values of $N$. This is thanks to the small amplitude of the tensor perturbations, generic in this class of models.
In the bottom panel, Figure~\ref{fig_8}, we note that for the exponential models, and depending on the number of $e$-folds, a single potential can generate values that fall outside the 95\%
confidence region (for $N=50$) and values that almost reach the  68\% confidence region for the Planck 2015 + BAO data (for $N=60$).
The model with potential $V=V_0/(2\pi \lambda V_0 T^2+1)$ lies precisely at
the centre of both confidence regions if $\lambda=0.0307$.

\subsection{An example from String Theory}
\label{subsec:example}
As mentioned earlier, a popular model of Tachyon Inflation is that of Eq.~\eqref{eq:sech}, with its observational parameters constrained in Figure~\ref{fig_8}. Now we show that our results may serve to constrain also specific models derived from string theory. As a particular example, we look at the action (with restored Natural units for clarity)\cite{Daniel},
\beq
\label{action:quevedo}
S = \int d^4x \sqrt{-g} \left[\frac{M_{\rm Pl}^{2}}{2} R -   \mathcal{A}\tilde{V}(\tilde{T})  \left( 1 +
\mathcal{B}g^{\mu \nu} \partial_{\mu} \tilde{T} \partial_{\nu} \tilde{T}  \right)^{1/2} \right], 
\eeq

\noindent  where the potential of Eq.~\eqref{eq:sech} written in terms of the fundamental string length $l_s = \sqrt{\alpha'}$:
\beq
\label{V:quevedo}
\tilde{V}(T) = \frac{\tilde{V}_0}{\cosh\left(\tilde{T} /\sqrt{2 \alpha'} \right)},
\eeq

The extra parameters $\mathcal{A}$ and $\mathcal{B}$ are dimensionless numbers coming from  the dimensional reduction of the action for the unstable brane system in string theory \cite{Daniel}. We note that our approach can constrain such numbers if we equate $V_0 = \mathcal{A}\tilde{V}_0$, $T =\sqrt{\mathcal{B}}\tilde{T}$ and our parameter $\lambda = \tilde{\lambda}/(\mathcal{AB})$, when we express \eqref{V:quevedo} in the form of \eqref{eq:sech} by writing,
\beq
\label{argument:quevedo}
x = \frac{\tilde{T}}{\sqrt{2 \alpha'}} = {\tilde{T}\sqrt{ \tilde{\lambda}\tilde{V}_0/M_{\rm Pl}^{2}} } = T\sqrt{ \lambda{V}_0/M_{\rm Pl}^{2} }.  
\eeq

\noindent After these substitutions, the model in \eqref{action:quevedo}, \eqref{V:quevedo} recovers the expressions in Eqs.~\eqref{eq:action} and \eqref{eq:sech}. Our fit to observables yields parameter values derived from the set $[V_0,\,\lambda,\,T(N=60)]  = [3.37\times 10^{-9}M_{\rm Pl}^{4}, 0.01, 2.12\times 10^{5}/M_{\rm Pl}]$, which consequently constrain the product
\beq
\alpha'\mathcal{B} = \frac{M_{\rm Pl}^{2}}{ 2 V_0\lambda} = 1.48\times 10^{10}/M_{\rm Pl}^{2}.
\eeq	

Thus a given choice of the fundamental string length $\alpha'$ is bound to generate a specific value of $\mathcal{B}$. In turn, this would set the value of the tachyon field of the string-inspired model with $\tilde{T}(N={60}) = T(N={60}) / \sqrt{\mathcal{B}}$. This would aid to reduce the value of the field trajectory during the inflationary stage; a desirable property to be explored in a follow-up paper \cite{barbosa2}. Further constraints to this particular model involve values for the radius of compactification (see \cite{kofman:linde,Daniel}), a subject to be explored elsewhere. For now, this example demonstrates that our analysis can constrain models motivated by high energy physics, with the bonus of a physical interpretation for our auxiliary parameter $\lambda$. 
\section{Discussion and Concluding Remarks}
\label{Sec:discuss}

We have studied the tachyon field inflationary scenario by means of the large-$N$ formalism.
We propose three universal classes of models for this scenario; namely the
perturbative, polynomial and exponential classes, presented in Section~\ref{unisec}. Such classification is naturally derived within the large-$N$ formalism and allows us to organize different models for the tachyon field inflation scenario in terms of their cosmological predictions, in the large-$N$ limit. A family of potentials is obtained for each one of the classes, with every realization corresponding to a numerical value of the parameter $\lambda$. We calculate for all the classes the form of the cosmological parameters in terms of the number of $e$-folds.

In the first case of study, the perturbative class, we recover the well known family of power-law potentials. We report on constraints to some of the most important realizations of the
model through up-to-date observations. This first class of models is particularly important since we can directly compare the models of Tachyon Scalar Field Inflation (TSFI) with those reconstructed for the Canonical Scalar Field Inflation (CSFI) scenario by means of their respective consistency relations, Eqs.~\eqref{TSFI:consistency} and \eqref{CSFI:consistency}, respectively. We find sensible differences in the results for the tensor-to-scalar ratio (see Eq.~\eqref{DCT}). We conclude that a precision of order one in $50$ in the detection of the tensor index with respect to a hypothetically observed value of the tensor-to scalar ratio is required in order to distinguish both scenarios.

For the second class of models, the polynomial class, we have derived a novel family of potentials within the tachyon inflationary scenario (see Eq.~\eqref{polynomial:potentials}).  The corresponding results in Figure~\ref{fig_FermionIII}(b) show that the polynomial class accommodates models that better fit the data, and mostly prefer a very small value of the tensor-to-scalar ratio.

For the exponential class, we focus on two specific cases for which
we can derive an explicit function of the potential $V(T)$. From both cases we find cases theoretically motivated in previous studies \cite{Sami:2002fs,Sen:2002an, tachyon1,cosh,Daniel}. The results obtained here complement previous analyses and provide a method to constrain those models.

We estimate higher order contributions of the inflationary perturbations by computing the non-Gaussianity parameter $f_{\rm NL}$ for the perturbations of the tachyon field model. In all cases this parameter in the local configuration approaches asymptotically to zero for large $N$, leading to a Gaussian distribution in the large-$N$ limit.

As argued in previous studies, high precision will be required in observations to distinguish tachyon inflation from its canonical counterpart. In particular, we note that the main differences arise for the cosmological parameters at second order in slow--roll. This is the main motivation to report the expressions of cosmological parameters at this order for each class. However, the difference between both scenarios at large $N$, where the featured formalism is valid, shows to be negligible.

The main feature of our study is the possibility of determining the observable parameters for several classes of potentials within the Tachyon inflationary scenario, by exploiting the large-$N$ formalism. The results in Figure \ref{fig_FermionIII} display important differences between the various realizations of the models in three classes of potentials. We conclude that the large-$N$ formalism is a powerful method to explore the richness of inflationary scenarios and to address their viability with enough precision for present and future observations.

\appendix
\section{The Fluid Approximation}
\label{app:A}
The fluid description of the tachyon field is valid as long as its evolution follows an adiabatic path. In the background such trajectory is derived from the first law of thermodynamics and implies the continuity equation. The stability of the adiabatic description is tested through perturbations and we show here that the non-adiabatic perturbations are suppressed in the inflationary stage.

\noindent The density and pressure of the tachyon field can be read from the matter action as %\cite{Garriga:1999vw}
\begin{equation}
\label{density:pressure}
\rho_T = \frac{V(T)}{\sqrt{1 - \dot T^2}}, \quad P_T = - {V(T)}{\sqrt{1 - \dot T^2}}.
\end{equation}
The adiabatic trajectory is one where the pressure and density change proportionally, according to the equation of state
\beq
w \equiv \frac{P}{\rho} =  - 1 + \dot{T}^2,
\eeq
which can be written in terms of the first Hubble flow parameter,
\begin{equation}
\label{eq:ESTATE}
\epsilon_{1}(N)=-\frac{\dot H}{H^2}=\frac{3}{2} \left(1+\frac{P}{\rho}\right)=\frac{3}{2} \left(1+\omega\right).
\end{equation}
It is because of this last equivalence that the first Hubble flow slow--roll parameter $\epsilon_1(N)$ has been dubbed the equation of state parameter. Yet the trajectories of the tachyon field may deviate from adiabaticity, which would break the fluid approximation for the tachyon perturbations. Here we show that this is not the case by looking at the adiabatic sound speed. This is the speed of perturbations along the adiabatic trajectory, and is given by
\beq
c_s^2 \equiv \frac{\dot{P_T}}{\dot\rho_T} = -  w \left(1 + \frac23 \frac{(\ln{V})'}{H\dot{T}}\right).
\eeq
This is easily derived from the continuity equation and the Hamilton-Jacobi Eq.~\eqref{H-J:2}. For matter perturbations, the departure from adiabaticity is denoted by the entropy perturbation $\delta P_{\rm nad}$:
\beq
\delta P_T = c_s^2 \delta \rho_T + \delta P_{\rm nad} = c_s^2 \delta \rho_T + P_T\Gamma.
\eeq
Here $\Gamma$, the entropy perturbation, parametrizes the difference between uniform density and uniform pressure slices. That is,
\beq
P_T \Gamma \equiv (c_{\rm eff}^2 - c_s^2)\delta\rho_{\rm com} = \frac23 \frac{(\ln{V})'}{H\dot{T}}\delta\rho_{\rm com}\,, \qquad \quad \mathrm{with}\quad c_{\rm eff} \equiv \frac{\partial P_T}{ \partial \dot{T}}  / \frac{\partial \rho_T}{ \partial \dot{T}} = w.
\eeq
Here we introduced the effective sound speed $c_{\rm eff}= w$, and note that the sub-index ${}_{\rm com}$ indicates that the matter density is measured in hypersurfaces orthogonal to world lines comoving with the fluid (comoving gauge). The comoving energy density perturbation is related to the gauge-invariant Bardeen metric potential $\Psi_B$ through the Poisson equation \cite{Bardeen:1980kt,Wands:2009ex},
\beq
4 \pi G a^2 \delta \rho_{\rm com} = \nabla^2 \Psi_B.
\eeq
a field equation valid even in a fluid with non-vanishing pressure \cite{Christopherson:2012kw}.
On the other hand, the difference in sound speeds can be written in terms of the slow--roll
parameters defined in \eqref{eq:defeps2},
\beq
c_{\rm eff}^2 - c_s^2 = \frac23  \frac{(\ln{V})'}{H\dot{T}} =  -  2 - \frac23 \frac{\epsilon_2}{(1 + \frac23 \epsilon_1)}.
\eeq
Combining the last two expressions we can write
\beq
\Gamma = - \frac{1}{w \rho_T} \left(2 + \frac23 \frac{\epsilon_2}{(1 + \frac23 \epsilon_1)}\right) \frac{\nabla^2 \Psi_B}{ 4 \pi G a^2} = \frac43 \left(1 + \frac13 \frac{\epsilon_2}{(1 + \frac23 \epsilon_1)}\right) \frac{\nabla^2 \Psi_B}{a^2 H^2}.
\eeq
The factor in large brackets is of order one in a slow--roll inflationary regime. However, inflationary perturbations are generated at horizon exit and subsequently expand to scales well above the Hubble horizon $r_H = 1 / H$, which implies that $\nabla^2 / (aH)^2 \ll 1$. The entropy perturbations are thus suppressed in the usual inflationary picture. This result agrees with that found in \cite{tachyon1} and justifies our assumption of adiabatic fluctuations when computing spectra and the bispectrum of the tachyon field.

\acknowledgments

We gratefully acknowledge support from \textit{Programa de Apoyo a Proyectos de Investigaci\'on e Innovaci\'on
Tecnol\'ogica} (PAPIIT) UNAM, IN103413-3,
% \textit{Teor\'ias de Kaluza-Klein, inflaci\'on y perturbaciones gravitacionales} 
and IA101414, 
%\textit{Fluctuaciones no-lineales en cosmolog\'{\i}a relativista} 
and SNI-CONACYT for support.  NBC and JDS acknowledge posdoctoral  grants from DGAPA-UNAM at ICF, and RRML a posdoctoral grant from CONACYT No. 647328.

%\paragraph{Note added.} This is also a good position for notes added
%after the paper has been written.


\end{document}